# Extended Fermi-Dirac and Bose-Einstein functions with applications to the family of zeta functions

by

**M. Aslam Chaudhry\*, Asghar Qadir\*\* and Asifa Tassaddiq\*\***

\* Department of Mathematics and Statistics
King Fahd University of Petroleum and Minerals
Dhahran 31261, Saudi Arabia
\*\* Center for Advanced Mathematics and Physics
National University of Science and Technology
H-12, Islamabad, Pakistan

maslam@kfupm.edu.sa, ashqadir@kfupm.edu.sa, asifashabbir@gmail.com

## Abstract

Fermi-Dirac and Bose-Einstein integral functions are of importance not only in quantum statistics but for their mathematical properties, in themselves. Here, we have extended these functions by introducing an extra parameter in a way that gives new insights into these functions and their relation to the family of zeta functions. These extensions are "dual" to each other in a sense that is explained. Some identities are proved for them and the relation between them and the general Hurwitz-Lerch zeta function $\Phi(z,s,v)$ is exploited to deduce new identities.

2000 Mathematics Subject Classification: Primary 11M06, 11M26, 11M36, 11M99, 42 A32, Secondary 11M06, 42A32

Keywords and phrases: Riemann zeta function, Hurwitz's zeta function, Hurwitz-Lerch zeta function, polylogarithm function, Fermi-Dirac integral function, Bose-Eienstein integral function

## I. Introduction

Most of the so called "special functions" arise as solutions of commonly occurring first or second order differential equations: the exponential functions for the first order and others like the Legendre Bessel, Laguerre, Hermite, etc. functions and polynomials. However, two of the most significant special functions do not come from differential equations: the gamma function and the zeta function (families). The gamma function lies at the base of all special functions as they are generally defined in terms of it. In some sense, it can be regarded as an "elementary function". The zeta function turns out to have significance well beyond its original purpose. It appears in quantum statistics as a special case of the Fermi-Dirac (FD) and Bose-Einstein (BE) integral functions. It also arises in quantum interference and the theory of quantum entanglement. In fact, it leads to a novel and more effective means of factorizing large numbers [3], [10], [14], [15]. There are many generalizations of the Riemann zeta function by Hurwitz, Lerch and others. Here we provide two extensions of the FD and BE functions. It turns out that in doing so, we find that we can "interpolate" between them in a way that follows the physical needs for doing so. Before going on to defining the new functions, it is worth putting the zeta function and the FD and BE functions in perspective for our purposes.

Riemann proved that the zeta-function

$$\zeta(s) := \sum_{n=1}^{\infty} \frac{1}{n^s} \qquad (s = \sigma + i\tau, \sigma > 1), \qquad (1.1)$$

has a meromorphic continuation to the complex plane. This satisfies the functional equation (see [20], p.13 (2.1.1))

$$\zeta(s) = 2(2\pi)^{-(1-s)} \cos(\frac{\pi}{2}(1-s))\Gamma(1-s)\zeta(1-s) := \chi(s)\zeta(1-s), \qquad (1.2)$$

and has simple zeros at $s = -2, -4, -6, \ldots$ called the *trivial zeros*. All the other zeros, called the *non-trivial zeros*, of the function are symmetric about the *critical line* $\sigma = 1/2$ in the *critical strip* $0 \leq \sigma \leq 1$. For the properties of the zeta function we refer to [1], [2], [5], [6], [8], [11], [12], [17], [16], [18], and [19]. The ratio $\chi(s)$ is defined in the sense of *Riemann's removable singularities theorem* ([2], p. 42) and it has the simple representation

$$\chi(s) = \frac{\pi^{s-\frac{1}{2}} \Gamma(\frac{1}{2} - \frac{s}{2})}{\Gamma(\frac{1}{2}s)} . \qquad (1.3)$$

Truesdell studied the properties of the polylogarithm (also known as de Jonquière's function)

$$Li_s(z) := \phi(z, s) := \sum_{n=1}^{\infty} \frac{z^n}{n^s} \qquad (z \in \xi), \qquad (1.4)$$

that generalizes the Riemann zeta function and has the integral representation

$$\phi(z, s) = \frac{z}{\Gamma(s)} \int_0^{\infty} \frac{t^{s-1}}{e^t - z} dt \qquad (|z| \leq 1 - \delta, \ \delta \in (0,1); z = 1, \ \sigma > 1). \qquad (1.5)$$

Note that "polylogarithm functions" should not be confused with "polylogarithm*ic* functions" nor with the "logarithmic integrals", all of which have similar notation. Further, if $z$ lies anywhere except on the segment of real axis from 1 to $\infty$, where a cut is imposed, the integral (1.5) defines an analytic function of $z$ for $\sigma > 0$. If $z = 1$, then the integral (1.5) coincides with the zeta function in $\sigma > 1$.

The polylogarithm function is further generalized to the Hurwitz-Lerch zeta function ([9], p. 27) by

$$\Phi(z,s,a) := \sum_{n=0}^{\infty} \frac{z^n}{(n+a)^s}$$

$(a \neq 0, -1, -2, -3, \ldots; s \in \mathbb{C}$ when $|z| < 1; \sigma > 1$ when $|z| = 1)$. (1.6)

This function has the integral representation ([9], p. 27(1.10)(3)),

$$\Phi(z,s,a) = \frac{1}{\Gamma(s)} \int_0^{\infty} \frac{t^{s-1} e^{-(a-1)t}}{e^t - z} dt$$

($\operatorname{Re}(a) > 0$, and either $|z| \leq 1, z \neq 1, \sigma > 0$ or $z = 1, \sigma > 1$). (1.7)

If a cut is made from 1 to $\infty$ along the positive real $z$-axis, $\Phi$ is an analytic function of $z$ in the cut $z$-plane provided that $\sigma > 0$ and $\operatorname{Re}(a) > 0$.

The Fermi-Dirac (FD) function ([7], p. 38(61))

$$F_{s-1}(x) := \frac{1}{\Gamma(s)} \int_0^{\infty} \frac{t^{s-1}}{e^{t-x} + 1} dt \qquad (s > 0),  \qquad (1.8)$$

and Bose Einstein (BE) function ([7], p. 449),

$$B_{s-1}(x) := \frac{1}{\Gamma(s)} \int_0^{\infty} \frac{t^{s-1}}{e^{t-x} - 1} dt \qquad (s > 1), \qquad (1.9)$$

are related to the polylogarithm-functions and the Hurwitz-Lerch function via

$$F_{s-1}(x) = -Li_s(-e^x) = -\phi(-e^x, s) = e^x \Phi(-e^x, s, 1) \quad (s > 0), \qquad (1.10)$$

$$B_{s-1}(x) = Li_s(e^x) = \phi(e^x, s) = e^x \Phi(e^x, s, 1) \qquad (s > 1). \qquad (1.11)$$

So many functions can be expressed in terms of the function $\Phi$. For example the polylogarithm function

$$Li_s(z) := \phi(z,s) := \sum_{n=1}^{\infty} \frac{z^n}{n^s} = z\Phi(z,s,1), \qquad (1.12)$$

Hurwitz's zeta function

$$\zeta(s,a) = \Phi(1,s,a) \qquad (1.13)$$

and the Riemann zeta function

$$\zeta(s) = \zeta(s,1) = \Phi(1,s,1), \qquad (1.14)$$

are special cases of the function $\Phi$. This shows that a new result, representation or relation proven for the function $\Phi$ will provide a wealth of results for the whole family of zeta functions. The proofs of the results satisfied by the Hurwitz-Lerch zeta function are cumbersome and do not reflect its simplicity. It will be seen that our extended FD and BE functions *do* provide relations that give simpler proofs of known results and some new results.

The plan of the paper is as follows: Some preliminaries are presented in the next section. We introduce classes of "good functions" in Section 3 and make use of Weyl's transform ([16], pp. 236-251) to find a series representation of these functions. Our extended functions are introduced in Sections 4 and 5 and some properties are investigated. We discuss the connection between these functions in Section 6. Some functional relations of these functions are discussed in Section 7. Using our general result of the series representation of our extended functions we derive known classical results for the Hurwitz-Lerch zeta, FD, BE and polylogarithm functions as special cases. The Riemann zeta function is recovered as a special case of these functions. Some properties of these functions have been obtained and relationships of these functions with the general Hurwitz-Lerch zeta function have been derived. An important feature of our approach is the desired simplicity of the proofs by using Weyl's fractional transform.

## 2. Preliminaries and the Mellin and Weyl transform representations

The classical Bernoulli and Euler polynomials of degree $n$ in $x$ are defined by ([18], (1.1), (1.2)),

$$\frac{te^{xt}}{e^t-1} = \sum_{n=0}^{\infty} B_n(x)\frac{t^n}{n!} \qquad (|t|<2\pi), \tag{2.1}$$

$$\frac{2e^{xt}}{e^t+1} = \sum_{n=0}^{\infty} E_n(x)\frac{t^n}{n!} \qquad (|t|<\pi). \tag{2.2}$$

The Bernoulli and Euler numbers are then defined by
$$B_n := B_n(0) \quad \text{and} \quad E_n := E_n(0) \qquad (n=0,1,2,3,...). \tag{2.3}$$

These polynomials and numbers are closely related to the zeta functions and we have (see [1], [5]),

$$\zeta(1-n,a) = -\frac{B(a)}{n} \qquad (n=1,2,3,...), \tag{2.4}$$

$$\zeta(2n) = \frac{(-1)^n(2\pi)^{2n}}{2(2n)!}B_{2n} \qquad (n=0,1,2,3,...). \tag{2.5}$$

The polynomials (2.1) and (2.2) are related via ([13], p. 29)

$$E_n(x) = \frac{2}{n+1}[B_{n+1}(x)+2^{n+1}B_{n+1}(\frac{x}{2})] \qquad (n=0,1,2,3,...). \tag{2.6}$$

The Mellin transform $\omega_M(s)$ ($s=\sigma+i\tau$) of a function $\omega(t)$ ($t>0$), if it exists as an analytic function in a strip $a<\sigma<b$ of analyticity, is defined by ([17], pp. 79-91)

$$\omega_M(s) := \int_0^{\infty} \omega(t)t^{s-1}dt \qquad (a<\sigma<b). \tag{2.7}$$

The Weyl transform $\Omega(s;x)$ ($s=\sigma+i\tau$, $x\geq 0$) of a function $\omega(t)$ ($t>0$), if it exists as an analytic function in $\sigma>0$, is defined by ([16], pp. 236-251)

$$\Omega(s;x) := W^{-s}[\omega(t)](x) := \frac{1}{\Gamma(s)}\int_0^{\infty}\omega(t+x)t^{s-1}dt = \frac{1}{\Gamma(s)}\int_x^{\infty}\omega(t)(t-x)^{s-1}dt$$
$$(\sigma>0,\ x\geq 0), \tag{2.8}$$

where, we define $\Omega(0;x)$ (provided $\omega(0)$ is also well defined) as,

$$\Omega(0;x) := \omega(x) \qquad (x\geq 0). \tag{2.9}$$

We have assumed $x\geq 0$. However, the function $\Omega(s;x)$ remains well defined for complex arguments provided $\text{Re}(x)\geq 0$. For $\sigma\leq 0$, the function $\Omega(s;x)$ is well defined for a class of "good functions" by ([16], p. 241)

$$\Omega(s;x) := (-1)^n \frac{d^n}{dx^n}[\Omega(n-s;x)] \qquad (x \geq 0), \qquad (2.10)$$

where, $n$ is the smallest non-negative integer greater or equal to $-\sigma := -\text{Re}(s)$. In particular, when $s = -m$ is a negative integer, we have ([16], p. 238(4.9)),

$$\Omega(-m;x) := (-1)^m \frac{d^m}{dx^m}[\Omega(0;x)] = (-1)^m \omega^{(m)}(x) \qquad (x \geq 0, \ m = 0,1,2,3,...). \qquad (2.11)$$

Since we have $W^{-(s+\beta)} = W^{-s}W^{-\beta}$ ([16], p. 242 (4.10)), this shows that

$$\Omega(s+\beta;x) := W^{-s}[\Omega(\beta;t)](x) := \frac{1}{\Gamma(s)}\int_0^\infty \Omega(\beta;t+x)t^{s-1}dt = \frac{1}{\Gamma(s)}\int_x^\infty \Omega(\beta;t)(t-x)^{s-1}dt$$

$$(\sigma > 0, \ \text{Re}(\beta) \geq 0). \qquad (2.12)$$

Moreover, it follows from (2.11) that

$$\Omega(0;x+t) = \Omega(0;x) + \sum_{n=1}^\infty \frac{(-1)^n}{n!}\Omega(-n;x)t^n \qquad (t \geq 0, \ x \geq 0). \qquad (2.13)$$

## 3. The classes $\wp(b)$ ($b > 0$) and $\wp(\infty)$ of "good functions" and a representation theorem

Following the terminology ([16], p. 237), we define classes of *good functions* as follows: If $\omega \in C^\infty(0,\infty)$ is integrable on any finite subinterval of $J := [0,\infty)$ and

$$|\omega(t)| = O(t^{-b}) \qquad (t \to \infty), \qquad (3.1)$$

then we say that $\omega \in \wp(b)$. If (3.1) is satisfied for every $b > 0$, then we say that $\omega \in \wp(\infty)$. Therefore, every $\omega \in \wp(b)$ satisfies (2.7) and

$$\Omega(s;x) := W^{-s}[\omega(t)](x) := \frac{1}{\Gamma(s)}\int_0^\infty \omega(t+x)t^{s-1}dt = \frac{1}{\Gamma(s)}\int_x^\infty \omega(t)(t-x)^{s-1}dt$$

$$(0 \leq \sigma < b, \ x \geq 0), \qquad (3.2)$$

whereas every $\omega \in \wp(\infty)$ satisfies (2.13).

**Theorem 3.1:** Let $\omega \in \wp(b)$ and $\Omega(s;x)$ ($x > 0$), be its Weyl transform. Then

$$\Omega(s;x) = \sum_{n=0}^\infty \frac{(-1)^n \Omega(s-n;0)x^n}{n!}$$

$$(0 \leq \sigma < b, x > 0). \qquad (3.3)$$

**Proof:** This follows from the fact that

$$\Omega(s-m;0) = (-1)^m \frac{d^m}{dx^m}[\Omega(s;x)]_{x=0} \qquad (x \geq 0, 0 < \sigma < b, \ m = 0,1,2,3,...). \qquad (3.4)$$

**Corollary 3.2:** Let $\omega \in \wp(b)$. Then

$$\omega(t) = \sum_{n=0}^\infty \frac{(-1)^n \Omega(-n;0)t^n}{n!} \qquad (0 \leq \sigma < b, t \geq 0). \qquad (3.5)$$

**Example 3.3:** We note that $\omega(t) = e^{-t} \in \wp(\infty)$ and $\Omega(s;0) \equiv 1$ that leads to the classical series representation

$$\omega(x) = e^{-x} = \sum_{n=0}^{\infty} \frac{(-1)^n x^n}{n!} \qquad (x \geq 0). \tag{3.6}$$

## 4. The extended Fermi-Dirac integral function $\Phi_\nu(s;x)$

In this section we introduce the first of our extended functions. An important aspect of the function is its transform representation and simplicity. It may be regarded as a natural extension of the FD function. In order to introduce the function, we consider

$$\varphi(t;\nu) := \frac{e^{-\nu t}}{e^t + 1} \qquad (\text{Re}(\nu) \geq 0,\ t \geq 0), \tag{4.1}$$

and note that $\varphi(t;\nu)$ ($\text{Re}(\nu) \geq 0$, $t \geq 0$) is integrable on any subinterval of $J := [0,\infty)$ and $\varphi(t;\nu) \in \wp(\infty)$. Therefore, the Weyl transform

$$\Phi_\nu(s;x) := \frac{1}{\Gamma(s)} \int_0^\infty t^{s-1} \varphi(t+x;\nu) dt := W^{-s}[\varphi(t;\nu)](x)$$

$$= \frac{1}{\Gamma(s)} \int_0^\infty t^{s-1} \varphi(t+x;\nu) dt = \frac{1}{\Gamma(s)} \int_x^\infty (t-x)^{s-1} \varphi(t;\nu) dt$$

$$(\sigma > 0,\ x \geq 0,\ \text{Re}(\nu) \geq 0), \tag{4.2}$$

is well defined. We call it the *extended FD function*. It is interesting to note that for $\nu = 0$ the function is related to FD integral function (2.7) via

$$\Phi_0(s;x) = F_{s-1}(-x) \qquad (\sigma > 0,\ x \geq 0). \tag{4.3}$$

For $x = 0$ in (4.2),

$$\Phi_\nu(s;0) := \frac{1}{\Gamma(s)} \int_0^\infty t^{s-1} \varphi(t;\nu) dt = \frac{1}{\Gamma(s)} \int_0^\infty \frac{e^{-\nu t} t^{s-1}}{e^t + 1} dt \qquad (\sigma > 0,\ \text{Re}(\nu) \geq 0). \tag{4.4}$$

The integral (4.4) remains absolutely convergent. Therefore, replacing the exponential function $e^{-\nu t}$ by its series representation and reversing the order of summation and integration, we find

$$\Phi_\nu(s;0) = \frac{1}{\Gamma(s)} \int_0^\infty t^{s-1} \varphi(t;\nu) dt = \sum_{n=0}^\infty \frac{(-1)^n \nu^n \Gamma(s+n)}{n! \Gamma(s)} \left( \frac{1}{\Gamma(s+n)} \int_0^\infty \frac{t^{s+n-1}}{e^t + 1} dt \right)$$

$$= \sum_{n=0}^\infty \frac{(-1)^n (s)_n \nu^n}{n!} \left( \frac{1}{\Gamma(s+n)} \int_0^\infty \frac{t^{s+n-1}}{e^t + 1} dt \right)$$

$$(\sigma > 0, \text{Re}(\nu) \geq 0). \tag{4.5}$$

However, the integral in (4.5) can be simplified in terms of zeta function ([5], p.301 (7.96)) to give

$$\frac{1}{\Gamma(s+n)} \int_0^\infty \frac{t^{s+n-1}}{e^t + 1} dt = (1 - 2^{1-s-n}) \zeta(s+n) \qquad (n = 0,1,2,3,\ldots,\ \sigma > 0). \tag{4.6}$$

From (4.5) and (4.6), we get
$$\Phi_\nu(s;0) = \sum_{n=0}^{\infty} \frac{(-1)^n (s)_n (1-2^{1-s-n})\zeta(s+n)\nu^n}{n!} \quad (\sigma > 0, \mathrm{Re}(\nu) \geq 0). \tag{4.7}$$

**Theorem 4.1:** The extended FD function can be expressed as an integral of itself as
$$\Phi_\nu(s+\beta;x) = \frac{1}{\Gamma(s)} \int_0^\infty \Phi_\nu(\beta;t+x) t^{s-1} dt = \frac{1}{\Gamma(\beta)} \int_0^\infty \Phi_\nu(s;t+x) t^{\beta-1} dt$$
$$(\sigma + \mathrm{Re}(\beta) > 0, x \geq 0, \mathrm{Re}(\nu) \geq 0). \tag{4.8}$$

**Proof:** This follows from the general representation (2.12).

**Corollary 4.2:** The Fermi-Dirac function can be expressed as an integral of itself as
$$F_{s+\beta-1}(x) = \frac{1}{\Gamma(s)} \int_0^\infty F_{\beta-1}(x+t) t^{s-1} dt$$
$$(\sigma + \mathrm{Re}(\beta) > 0, x \leq 0). \tag{4.9}$$

**Proof:** This follows from (4.8) when $\nu = 0$.

**Remark 4.3:** Note that the extended FD function reduces to the series representation (4.7) when $x = 0$. Hence, for $0 \leq \mathrm{Re}(\nu) < 1$ it is an entire function. In particular for $\nu = 0$ and $x = 0$, we have the entire function
$$\Phi_0(s;0) = (1 - 2^{1-s})\zeta(s) \quad (\sigma > 0). \tag{4.10}$$

Further, as was shown in (4.3), $\Phi_0(s;x) = F_{s-1}(-x)$, which is why we say that it is a natural generalization of the FD function.

Putting the value of $\varphi(t;\nu)$ in (4.4), we obtain
$$\Phi_\nu(s;x) = \frac{e^{-(\nu+1)x}}{\Gamma(s)} \int_0^\infty \frac{e^{-\nu t}}{e^t + e^{-x}} t^{s-1} dt \quad (\sigma > 0, x \geq 0, \mathrm{Re}(\nu) \geq 0), \tag{4.11}$$

which shows that the extended FD function is related to Hurwitz-Lerch zeta function via
$$\Phi_\nu(s;x) = e^{-(\nu+1)x} \Phi(-e^{-x}, s, \nu+1) \quad (\sigma > 0, x \geq 0, \mathrm{Re}(\nu) \geq 0). \tag{4.12}$$

Therefore, we have the series representation
$$\Phi_\nu(s;x) = e^{-(\nu+1)x} \sum_{n=0}^{\infty} \frac{(-1)^n e^{-nx}}{(n+\nu+1)^s}$$
$$(\sigma > 0, x \geq 0, \mathrm{Re}(\nu) \geq 0). \tag{4.13}$$

**Theorem 4.4:** The extended FD function has the power series representation
$$\Phi_\nu(s;x) = \Phi_\nu(s;0) + \sum_{n=1}^{\infty} \frac{(-1)^n \Phi_\nu(s-n;0) x^n}{n!} \quad (\sigma > 0, x \geq 0, \mathrm{Re}(\nu) \geq 0). \tag{4.14}$$

**Proof:** It has the Weyl transform representation (4.4). Therefore, application of the general result (3.3) leads to the proof of (4.14).

**Corollary 4.5:** The FD function has the power series representation

$$F_{s-1}(x) = F_{s-1}(0) + \sum_{n=1}^{\infty} \frac{F_{s-n-1}(0)x^n}{n!}$$

$$= (1-2^{1-s})\zeta(s) + \sum_{n=1}^{\infty} \frac{(1-2^{1-s-n})\zeta(s-n)x^n}{n!}$$

$$(\sigma > 0,\ x \geq 0). \tag{4.15}$$

**Proof:** This follows from (4.14) on replacing $x$ by $-x$, putting $v = 0$ and using

$$\Phi_0(s;0) = F_{s-1}(0) = (1-2^{1-s})\zeta(s) \qquad (\sigma > 0). \tag{4.16}$$

## 5. The extended Bose-Einstein integral function

In this section we introduce our second extended function which also has the transform representation. The function may be regarded as a natural extension of the BE function. In order to introduce this function, we consider

$$\psi(t;v) := \frac{e^{-vt}}{e^t - 1} \qquad (\mathrm{Re}(v) \geq 0,\ t \geq 0). \tag{5.1}$$

Of course, the function (5.1) does not have a convergent Mellin transform when $0 < \sigma < 1$. We define

$$\Psi_v(s;x) := \frac{1}{\Gamma(s)} \int_0^{\infty} t^{s-1} \psi(t+x;v) dt := W^{-s}[\psi(t;v)](x)$$

$$= \frac{1}{\Gamma(s)} \int_0^{\infty} t^{s-1} \psi(t+x;v) dt = \frac{1}{\Gamma(s)} \int_x^{\infty} (t-x)^{s-1} \psi(t;v) dt$$

$$(\sigma > 1,\ x \geq 0,\ \mathrm{Re}(v) \geq 0). \tag{5.2}$$

Note that for $v = 0$ it is related to the BE function via

$$\Psi_0(s;x) = B_{s-1}(-x) \qquad (\sigma > 1,\ x \geq 0). \tag{5.3}$$

Putting the value of $\psi(t;v)$ in (5.2) yields

$$\Psi_v(s;x) = \frac{e^{-(v+1)x}}{\Gamma(s)} \int_0^{\infty} \frac{e^{-vt}}{e^t - e^{-x}} t^{s-1} dt$$

$$(\sigma > 1,\ x = 0,\ \mathrm{Re}(v) \geq 0\ ; x > 0, \sigma > 0). \tag{5.4}$$

Hence the extended BE function is related to Hurwitz's zeta function via

$$\Psi_v(s;x) = e^{-(v+1)x}\Phi(e^{-x},s,v+1)$$

$$(\sigma > 1,\ x = 0, \mathrm{Re}(v) \geq 0\ ; x > 0, \sigma > 0). \tag{5.5}$$

Therefore, we have the series representation

$$\Psi_v(s;x) = e^{-(v+1)x} \sum_{n=0}^{\infty} \frac{e^{-nx}}{(n+v+1)^s} \qquad (\sigma > 1,\ x \geq 0,\ \mathrm{Re}(v) \geq 0). \tag{5.6}$$

In particular, we have

$$\Psi_v(s;0) = \zeta(s,v+1) \qquad (\sigma > 1,\ \mathrm{Re}(v) \geq 0), \tag{5.7}$$

which shows that $\Psi_\nu(s;0)$ has a meromorphic continuation ([1], p. 254) to the complex plane that leads to the series representation

$$\Psi_\nu(s;0) = \sum_{n=0}^{\infty} \frac{(-1)^n (s)_n \zeta(s+n) v^n}{n!} \qquad (\sigma > 0, \sigma \neq 1, 0 \leq v < 1). \qquad (5.8)$$

From (5.7) we also see that $\Psi_\nu(s;0)$ can be analytically continued in $s$ to the complex half-plane $\sigma \leq 0$ ([1], p. 254). In particular, ([1], p. 264, p. 275)

$$\Psi_\nu(-n;0) = -\frac{B_{n+1}(v+1)}{n+1} \qquad (n = 0,1,2,3,\ldots), \qquad (5.9)$$

where $B_n(x)$ ($n = 0,1,2,\ldots$) are the Bernoulli polynomials ([1], p. 263).

In fact, the relation (5.5) shows that the extended BE function also has a meromorphic continuation to the entire complex plane.

**Theorem 5.1:**

$$\Psi_a(s;x) = q^{-s} e^{\frac{ax(1-q)}{q}} \sum_{j=1}^{q} e^{\frac{xj(1-q)}{q}} \Psi_{\frac{a+j-q}{q}}(s;qx) \qquad (\sigma > 1, x \geq 0, \mathrm{Re}(a) \geq 0). \qquad (5.10)$$

**Proof:** The Hurwitz-Lerch zeta function has representation ([18], (3.8))

$$\Phi(z,s,a) = q^{-s} \sum_{j=1}^{q} \Phi\left(z^q, s, \frac{a+j-1}{q}\right) z^{j-1} \qquad (5.11)$$

and hence (5.10) follows from (5.5) and (5.11).

**Corollary 5.2:** $\Psi_a(s;0) = q^{-s} \sum_{j=1}^{q} \Psi_{\frac{a+j-q}{q}}(s;0) \qquad (\sigma > 1, \mathrm{Re}(a) \geq 0). \qquad (5.12)$

**Proof:** This follows from (5.10) when $x = 0$.

**Remark 5.3:** Relation (5.12) can be rewritten in terms of the Hurwitz's zeta function

$$\zeta(s, a+1) = q^{-s} \sum_{j=1}^{q} \zeta(s, \frac{a+j}{q}) \qquad (q = 1,2,3,\ldots), \qquad (5.13)$$

which is a well known identity. In particular, for $a = 0$, we get another well know result

$$\zeta(s) = \zeta(s,1) = q^{-s} \sum_{j=1}^{q} \zeta(s, \frac{j}{q}) \qquad (q = 1,2,3,\ldots). \qquad (5.14)$$

With $q = 2$ in (5.14), we recover the identity ([20], p. 36)

$$\zeta(s)(2^s - 1) = \zeta(s, \frac{1}{2}). \qquad (5.14)$$

# 6. Connection between the two extended functions

The two extended functions are simply related to each other, as we show in this section. We also deduce a connection between the FD and BE functions as a special case.

**Theorem 6.1:** The FD and BE extended functions are related by
$$\Phi_{2\nu}(s;x) = \Psi_{2\nu}(s;x) - 2^{1-s}\Psi_{\nu}(s,2x) \qquad (x \geq 0,\ \sigma > 0,\ \text{Re}(\nu) \geq 0). \tag{6.1}$$

**Proof:** Since,
$$\frac{e^{-2\nu t}}{e^{2t}-1} = \frac{1}{2}\left(\frac{e^{-2\nu t}}{e^{t}-1} - \frac{e^{-2\nu t}}{e^{t}+1}\right), \tag{6.2}$$

from (4.1) and (5.1) we have
$$\psi(2t;\nu) = \frac{1}{2}[\psi(t;2\nu) - \varphi(t;2\nu)]. \tag{6.3}$$

Taking the Weyl transform of both sides in (6.3) and using

$$W^{-s}[\psi(2t;\nu)](x) = 2^{-s}\Psi_{\nu}(s;2x), \tag{6.4}$$

we obtain

$$2^{-s}\Psi_{\nu}(s;2x) = \frac{1}{2}[\Psi_{2\nu}(s;x) - \Phi_{2\nu}(s;x)], \tag{6.5}$$

which is the transposed form of (6.1).

**Corollary 6.2:** The FD and BE functions are related via
$$F_{s-1}(x) = B_{s-1}(x) - 2^{1-s}B_{s-1}(2x) \qquad (x \geq 0,\ \sigma > 0). \tag{6.6}$$

**Proof:** This follows from (6.1) when we put $\nu = 0$ and replace $x$ by $-x$.

**Theorem 6.3:** The extended FD and BE functions are also related via
$$\Psi_{\nu}(s;x) = e^{-i(\nu+1)\pi}\Phi_{\nu}(s;x+\pi i) \qquad (x \geq 0,\ \sigma > 0,\ \text{Re}(\nu) \geq 0). \tag{6.7}$$

**Proof:** Replacing $x$ by $x + \pi i$ in (4.2), we find

$$\Phi_{\nu}(s;x) := \frac{1}{\Gamma(s)}\int_{0}^{\infty} t^{s-1}\varphi(t+x+\pi i;\nu)dt \qquad (x \geq 0,\ \sigma > 0,\ \text{Re}(\nu) \geq 0). \tag{6.8}$$

However, we have
$$\varphi(t+x+\pi i;\nu) = \frac{e^{-\pi\nu i}e^{-\nu(t+x)}}{-e^{t+x}+1} = -\frac{e^{-\nu\pi i}e^{-\nu(t+x)}}{e^{t+x}-1} = \frac{e^{-(\nu+1)\pi i}e^{-\nu(t+x)}}{e^{t+x}-1}. \tag{6.9}$$
$$= e^{-(\nu+1)\pi i}\psi(t+x;\nu)$$

By taking the Mellin transform of both sides in (6.9) we obtain the result.

Note that the extended FD and BE functions are dual to each other in the sense that (6.7) can easily be inverted. This fact is of relevance for providing a function for anyons [4] corresponding to the Fermi-Dirac and Bose-Einstein integral functions.

**Theorem 6.4:**
$$\Phi_{\nu+1}(s,x) = 2^{-s}[\Psi_{\frac{\nu}{2}}(s;2x) - \Psi_{\frac{\nu+1}{2}}(s;2x)]$$
$$(\sigma > 0, \text{Re}(\nu) > 0 ; \text{ for } \text{Re}(\nu) = 0 \text{ (but } \text{Im}(\nu) \neq 0), 0 < \sigma < 1). \quad (6.10)$$

**Proof:** Replacing $\Psi_a(s;2x)$ by its integral representation, we obtain

$$\Psi_{\frac{\nu}{2}}(s;2x) - \Psi_{\frac{\nu+1}{2}}(s;2x) = \frac{1}{\Gamma(s)} \int_{2x}^{\infty} \frac{(e^{-\frac{\nu t}{2}} - e^{-\frac{(\nu+1)t}{2}})(t-2x)^{s-1}}{e^t - 1} dt$$

$$= \frac{1}{\Gamma(s)} \int_{2x}^{\infty} \frac{e^{-\frac{(\nu+1)t}{2}}(e^{\frac{t}{2}}-1)(t-2x)^{s-1}}{(e^{\frac{t}{2}}-1)(e^{\frac{t}{2}}+1)} dt = \frac{1}{\Gamma(s)} \int_{2x}^{\infty} \frac{e^{-\frac{(\nu+1)t}{2}}(t-2x)^{s-1}}{(e^{\frac{t}{2}}+1)} dt$$

$$(x \geq 0, \sigma > 0, \text{Re}(\nu) > 0 ; \text{ for } \text{Re}(\nu) = 0 \text{ (but } \text{Im}(\nu) \neq 0), 0 < \sigma < 1). \quad (6.11)$$

The transformation $t = 2\tau$ in (6.11) leads to

$$\Psi_{\frac{\nu}{2}}(s;2x) - \Psi_{\frac{\nu+1}{2}}(s;2x) = \frac{2^s}{\Gamma(s)} \int_x^{\infty} \frac{e^{-(\nu+1)\tau}(\tau-x)^{s-1}}{e^{\tau}+1} d\tau,$$

which is the transposed form of (6.10).

**Corollary 6.5:** $\Phi_\nu(s,0) = 2^{-s}[\zeta(s;\frac{\nu}{2}) - \zeta(s;\frac{\nu+1}{2})]$
$$(\sigma > 0, \text{Re}(\nu) > 0 ; \text{ for } \text{Re}(\nu) = 0 \text{ (but } \text{Im}(\nu) \neq 0), 0 < \sigma < 1). \quad (6.12)$$

**Proof:** Using $\Psi_a(s;0) = \zeta(s, a+1)$ in (6.11), we arrive at (6.12).

## 7. Functional relations for the extended FD function

Functional relations and difference equations are important for the study of special functions. For example, the Bernoulli polynomials satisfy the difference equation ([1], p. 265(18))

$$B_n(x+1) - B_n(x) = nx^{n-1} \qquad (n \geq 1). \quad (7.1)$$

One would like to know if our extended functions also satisfy such relations. It is found that this is the case.

**Theorem 7.1:** The extended FD function $\Phi_\nu(s;x)$ satisfies the difference equation

$$\Phi_{\nu+1}(s;x) + \Phi_\nu(s;x) = (\nu+1)^{-s} e^{-(\nu+1)x} \qquad (x \geq 0, \nu \geq 0). \quad (7.2)$$

**Proof:** We have the identity

$$\frac{e^{-(\nu+1)t}}{e^t+1} + \frac{e^{-\nu t}}{e^t+1} = e^{-(\nu+1)t} \quad , \quad (7.3)$$

that can be rewritten as

$$\varphi(t;\nu+1) + \varphi(t;\nu) = e^{-(\nu+1)t}. \quad (7.4)$$

However, we have

$$W[e^{-(\nu+1)t}](x) = (\nu+1)^{-s} e^{-(\nu+1)x} . \tag{7.5}$$

Applying the Weyl transform on both sides in (7.4) and using (7.5), we arrive at (7.2).

**Corollary 7.2:** $\quad \Phi_\nu(s;0) + \Phi_{\nu-1}(s;0) = \nu^{-s} \qquad (x \geq 0, \nu \geq 1). \tag{7.6}$

**Proof:** This follows from (7.2) by putting $x = 0$ and replacing $\nu$ by $\nu - 1$.

**Corollary 7.3:** The Hurwitz-Lerch zeta function satisfies the difference equation
$$\Phi(z,s,\nu) - z\Phi(z,s,\nu+1) = \nu^{-s} \qquad (\nu \geq 1). \tag{7.7}$$
**Proof:** This follows from (4.13) and (7.2).

**Remark 7.4:** The extended FD function and the Bernoulli polynomials are related via
$$\Phi_\nu(-n;\pi i) = e^{-i\pi\nu} \frac{B_{n+1}(\nu+1)}{n+1} \qquad (n=0,1,2,3,...). \tag{7.8}$$
Putting $s = -n$ and $x = \pi i$ in (7.2) and using (7.9), we find the classical relation
$$\frac{B_{n+1}(\nu+1)}{n+1} - \frac{B_{n+1}(\nu)}{n+1} = \nu^n \qquad (n \geq 0), \tag{7.9}$$
between the Bernoulli polynomials. This shows that the difference equation (7.2) that can be rewritten as
$$\Phi_\nu(s;x) + \Phi_{\nu-1}(s;x) = \nu^{-s} e^{-\nu x} \qquad (x \geq 0, \nu \geq 1), \tag{7.10}$$
is the most general form of the difference equation satisfied by the family of the zeta functions. Similarly, putting $z = 1$ in (7.7) and using $\zeta(s,a) = \Phi(1,s,a)$, we recover the difference equation
$$\zeta(s,\nu) - \zeta(s,\nu+1) = \nu^{-s} \qquad (\nu \geq 1), \tag{7.11}$$
satisfied by the Hurwitz's zeta function.

## 8. Concluding remarks and discussion

We have provided an extension of the FD and BE integral functions. These functions are closely related to the family of zeta functions. By this extension their connection becomes more manifest and can be explored further. We have seen some of the connections here. There are bound to be many others, some of which might prove fruitful to explore.

The FD and BE functions arose in the distribution functions for quantum statistics. While the Maxwell distribution provides the velocity distribution of molecules of a classical gas, the FD and BE functions come from the velocity distribution of a quantum gas. The crucial difference is that the former are (in principle) distinguishable particles while the latter are, even in principle, *indistinguishable*. According to the standard theory, all particles either have spins of integer or half integer multiples of Planck's constant (divided by $2\pi$), $\hbar$. If the spin is half integer they are called Fermions and their velocity distribution is given by the FD function. If it is integer they are called Bosons and their velocity distribution is given by the BE functions. Some phenomena were discovered in which the particles behaved as if they were indistinguishable but neither fermions nor bosons. They were called *anyons*. The extensions may help to describe them [4].

It is worth noting that the extensions have led to a relation between the well known FD and BE integral functions that had not been seen before. It also highlighted the close relationship between these special functions and the zeta functions. Whereas the usual "special functions of mathematical physics" arise from second order partial differential equations, when solving by means of separation of variables, leading to Sturm-Liouville problems, the zeta function was not seen to be related to physical problems but was thought to turn up only in number theory. More recently it has been found to arise in the theory of quantum entanglement. This may have seemed surprising. However, the close connection of the zeta with the FD and BE functions brought out by the extension makes this new appearance more understandable.

**Acknowledgements** Two of the authors (MAC and AQ) are grateful to the King Fahd University of Petroleum and Minerals for providing excellent research facilities. AT acknowledges her indebtedness to the Higher Education Commission of Pakistan for the Indigenous PhD Fellowship.